\documentclass{article}
\usepackage{spconf,amsmath,graphicx}
\usepackage{makecell}
\usepackage{threeparttable}

\title{Shaking Acoustic Spectral Sub-bands Can Better Regularize Learning in Affective Computing}
\name{Che-Wei Huang, Shrikanth Narayanan}
\address{University of Southern California, Los Angeles, CA 90089 \\ cheweihu@usc.edu, shri@sipi.usc.edu}
\begin{document}
%
\maketitle
\begin{abstract}
In this work, we investigate a recently proposed regularization technique based on multi-branch architectures, called Shake-Shake regularization, for the task of speech emotion recognition. In addition, we also propose variants to incorporate domain knowledge into model configurations. The experimental results demonstrate: $1)$ independently shaking sub-bands delivers favorable models compared to shaking the entire spectral-temporal feature maps.
$2)$ with proper patience in early stopping, the proposed models can simultaneously outperform the baseline and maintain a smaller performance gap between training and validation.

\end{abstract}
\section{Introduction}
\label{sec:intro}
Deep convolutional neural networks have been successfully applied to several pattern recognition tasks such as image recognition \cite{He_2016_CVPR}, machine translation \cite{gehring2017} and speech emotion recognition \cite{Huang2017}. Currently, to successfully train a deep neural network, one needs either a sufficient number of training samples to implicitly regularize the learning process, or employ techniques like weight decay and dropout \cite{Srivastava:2014} and its variants to explicitly keep the model from over-fitting. 

In the recent years, one of the most popular and successful architectures is the residual neural network (ResNet) \cite{He_2016_CVPR}. The ResNet architecture was designed based on a key assumption that it is more efficient to optimize the residual term than the original task mapping. Since then, a great deal of effort in machine learning and computer vision has been dedicated to study the multi-branch architecture.
    
Deep convolutional neural networks have also gained much attention in the community of affective computing mainly because of its outstanding ability to formulate discriminative features for the top-layer classifier. Usually the number of parameters in a model is far more than the number of training samples and thus it requires heavy regularization to train deep neural networks for affective  computing. However, since the introduction of batch normalization \cite{batchnorm}, the gains obtained by using dropout for regularization have decreased \cite{batchnorm,Zagoruyko2016WRN,Huang2016}. Yet, multi-branch architectures have emerged as a promising alternative.

Regularization techniques based on multi-branch architectures such as Shakeout \cite{KangLT16} and Shake-Shake \cite{gastaldi2017} have delivered impressive performances on standard image datasets such as the CIFAR-10 \cite{cifar10}. In a clever way, both of them utilize multiple branches to learn different aspects of the relevant information and then a summation in the end follows for information alignment among branches. Instead of using multiple branches, a recent work \cite{randomproj} based on a mixture of experts showed that randomly projecting samples is able to break the structure of adversarial noise that could easily confound the model and as a result mislead the learning process. Despite not being an end-to-end approach, it shares the same idea of integrating multiple streams of model-based diversity.  

In this work, we study the Shake-Shake regularized ResNet for speech emotion recognition. In addition to shaking the entire spectral-temporal feature maps with the same strength, we propose to address different spectral sub-bands independently based on our hypothesis of the non-uniform distribution of affective information over the spectral axis. There has been work on multi-stream framework in speech processing. For example, Mallidi et al. \cite{mallidi2016} designed a robust speech recognition system using multiple streams, each of them attending to a different part of the feature space, to fight against noise. However, lacking both multiple branches and the final information alignment, the design philosophy is fundamentally different from that of multi-branch architectures. In fact, we intend to serve this work as a bridge between the multi-stream framework and the multi-branch architecture. 

\subsection{Shake-Shake Regularization}
\label{subsec:shakeshake_regular}
Shake-Shake regularization \cite{gastaldi2017} is a recently proposed technique to regularize training of deep convolutional neural networks for image recognition tasks. This regularization technique based on multi-branch architectures promotes stochastic mixtures of forward and backward propagations from network branches in order to create a flow of model-based adversarial learning samples/gradients during the training phase. 
Owing to it excellent ability to combat over-fitting even in the presence of batch normalization, the Shake-Shake regularized $3$-branch residual neural network \cite{gastaldi2017} has achieved the current state-of-the-art performance on the CIFAR-10 image dataset.

\begin{figure}[t]
  \centering
  \includegraphics[width=0.45\textwidth,height=4cm]{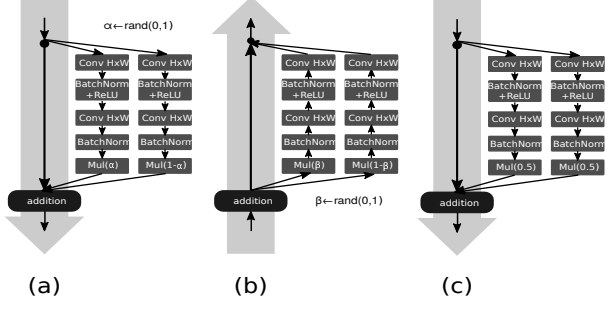}
  \caption{An overview of a $3$-branch shake-shake regularized residual block \cite{gastaldi2017}. (a) Forward propagation during the training phase (b) Backward propagation during the training phase (c) Testing phase. The coefficients $\alpha$ and $\beta$ are sampled from the uniform distribution over $[0,1]$ to scale down the forward and backward flows during the training phase.}
\label{fig:shakeshake}
\end{figure}

An overview of a $3$-branch shake-shake regularized ResNet is depicted in Fig. \ref{fig:shakeshake}. In addition to the short-cut flow (in light gray), there are other two residual branches $\mathbf{B}(x)$, each of them consisting of a sequence of layers stacked in order: Conv$H\times W$, Batch Normalization, ReLU, Conv$H\times W$, Batch Normalization, where Conv$H\times W$ represents a convolutional layer with filters of size $H\times W$ without pooling and ReLU is the rectified linear unit $\rm{ReLU}(x) = \max(0,x)$. 


The Shake-Shake regularization adds to the aggregate of the output of each branch an additional layer, called the ShakeShake layer, to randomly generate adversarial flows in the following way:
\begin{equation}
\mathbf{\rm{ShakeResNet}}_N(\mathbf{X}) = \mathbf{X} + \sum_{n=1}^{N} \text{ShakeShake}\left(\left\{\mathbf{B}_n(\mathbf{X})\right\}_{n=1}^{N}\right)
\nonumber
\end{equation}
where in the forward propagation for $\mathbf{a} = [\alpha_1, \cdots, \alpha_N]$ sampled from the ($N$$-$$1$)-simplex (Fig. \ref{fig:shakeshake} (a))
\begin{equation}
\mathbf{\text{ShakeResNet}}_N(\mathbf{X}) = \mathbf{X} + \sum_{n=1}^N \alpha_n \mathbf{B}_n(\mathbf{X}),
\nonumber
\end{equation}
while in the backward propagation for $\mathbf{b} = [\beta_1, \cdots, \beta_N]$ sampled from the ($N$$-$$1$)-simplex and $\mathbf{g}$ the gradient from the top layer, the gradient entering into $\mathbf{B}_n(x)$ is $\beta_n\mathbf{g}$ (Fig. \ref{fig:shakeshake} (b)).
At the testing time, the expected model is then evaluated for inference by taking the expectation of the random sources in the architecture (Fig. \ref{fig:shakeshake} (c)).  

In each mini-batch, to apply scaling coefficients $\alpha$ or $\beta$ either on the entire mini-batch or on each individual sample independently can also make a difference \cite{gastaldi2017}. 

\section{Proposed Models}
\label{sec:proposed}
In addition to batch- or sample-wise shaking, when it comes to the area of acoustic processing, there is another orthogonal dimension to consider: the spectral domain. Leveraging domain knowledge, our proposed models are based on a simple but plausible hypothesis that affective information is distributed non-uniformly over the spectral axis \cite{Lee04emotionrecognition}. Therefore, there is no reason to enforce the entire spectral axis to be shaken with the same strength concurrently. Furthermore, adversarial noise may exist and extend over the spectral axis. By deliberately shaking spectral sub-bands independently, the structure of adversarial noise may be broken and become less confounding to the model.  

\begin{figure}[th]
  \centering
  \includegraphics[width=0.4\textwidth,height=3cm]{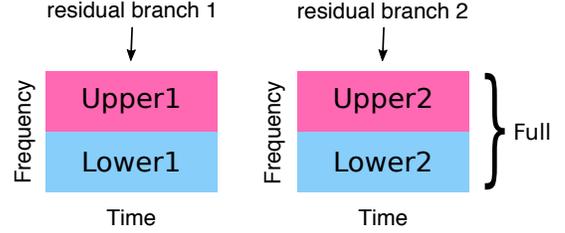}
  \caption{An illustration for the sub-band definitions.}
\label{fig:shakesubband}
\end{figure}

Before we formally define the proposed models, we introduce the definition of sub-bands first. Fig. \ref{fig:shakesubband} depicts the definition for sub-bands in a $3$-branch residual block. Here we slightly abuse the notations of frequency and time because after two convolutional layers these axes are not exactly the same as those of input to the branches; however, since convolution is a local operation they still hold the corresponding spectral and temporal nature. At the output of each branch, we define the high-frequency half to be the upper sub-band while the low-frequency half to be the lower sub-band. We take the middle point on the spectral axis to be the border line for simplicity. The entire output is called the full band.

Having defined these concepts, we denote $\mathbf{X}$ the input to a residual block, $\mathbf{X}^i$ the full band from the $i$-th branch, $\mathbf{X}^i_u$ the upper sub-band from the $i$-th branch and $\mathbf{X}^i_l$ the lower sub-band from the $i$-th branch. Naturally, the relationship between them is given by $\mathbf{X}^i = \left[\mathbf{X}^i_u | \mathbf{X}^i_l\right]$. We also denote $\mathbf{Y}$ the output of a Shake-Shake regularized residual block.

To demonstrate that shaking sub-bands can better regularize the learning process, we propose the following models for benchmarking:
\begin{enumerate}
\item Shake the full band (\textbf{Full})
\begin{equation}
\mathbf{Y} = \mathbf{X} + \sum_{n=1}^{N} \text{ShakeShake}\left(\left\{\mathbf{X}^n\right\}_{n=1}^{N}\right).
\end{equation}
\item Shake the upper sub-band (\textbf{Upper})
\begin{eqnarray}
\mathbf{Y} &=& \mathbf{X} \\
&+& 
\left[
\sum_{n=1}^{N} \mathrm{ShakeShake}
\left(\left\{\mathbf{X}^n_u\right\}_{n=1}^{N}\right)
 \middle|
\sum_{n=1}^N\mathbf{X}^n_l
\right].
\nonumber
\end{eqnarray}
\item Shake the lower sub-band (\textbf{Lower})
\begin{eqnarray}
\mathbf{Y} &=& \mathbf{X} \\
&+& 
\left[
\sum_{n=1}^{N} \mathbf{X}^n_u
 \middle|
\sum_{n=1}^N\text{ShakeShake} 
\left( \left\{ 
\mathbf{X}^n_l
\right\}_{n=1}^{N}\right)
\right].
\nonumber
\end{eqnarray}
\item Shake both sub-bands but independently (\textbf{Both})
\begin{eqnarray}
\mathbf{Y} &=& \mathbf{X} + \left[ \mathbf{Y}_u |\mathbf{Y}_l\right],\\
\mathbf{Y}_u &=& \sum_{n=1}^{N} \text{ShakeShake}\left(\left\{\mathbf{X}^n_u\right\}_{n=1}^{N}\right), \nonumber\\
\mathbf{Y}_l &=& \sum_{n=1}^N\text{ShakeShake} \left( \left\{ \mathbf{X}^n_l\right\}_{n=1}^{N}\right). \nonumber
\end{eqnarray}
\end{enumerate} 

\section{Experiments}
\label{sec:experiments}

\subsection{Experiment I}
\subsubsection{Datasets}
\label{subsec:datasets}
We use four publicly available emotion corpora to demonstrate the effectiveness of the proposed models, including the Ryerson Audio-Visual Database of Emotional Speech and Song (RAVDESS) \cite{ravdess}, the eNTERFACE'05 Audio-Visual Emotion Database \cite{enterface05}, the EMOVO Corpus \cite{emovo} and the Surrey Audio-Visual Expressed Emotion (SAVEE) \cite{savee}. All of these corpora are multi-modal in which speech, facial expression and text all convey a certain degree of affective information. However, in this paper we solely focus on the acoustic modality for experiments.


The intersection of emotional classes in these four corpora consists of joy, anger, sadness and fear. Therefore, we formulate the experimental task into a sequence classification of $4$ classes. In particular, we employ both speaking and singing sets from the RAVDESS corpus, all of the eNTERFACE'05, EMOVO and SAVEE corpora. However, one of the female actors in RAVDESS corpus does not have the singing part and we thus leave her speech part out of the experiments as well. The actor $23$ in eNTERFACE'05 has only $3$ utterances portraying joy, which makes the emotional class distribution slightly imbalanced. As a result, we have $2885$ utterances in total. Table \ref{tab:datasets} summarizes the information about these four corpora. 

\begin{table}[ht]
  \centering
  \begin{tabular}{lccccc}
  \Xhline{4\arrayrulewidth}
  Corpus & No. & \multicolumn{4}{c}{No. Utterances}\\
  \cline{3-6}
  & Actors & joy & anger & sadness & fear \\
  \Xhline{2\arrayrulewidth} 
  RAVDESS           & $23$ & $368$ & $368$ & $368$ & $368$\\
  eNTERFACE         & $42$ & $207$ & $210$ & $210$ & $210$ \\ 
  EMOVO              & $6$  & $84$ & $84$ & $84$ & $84$\\
  SAVEE              & $4$  & $60$ & $60$ & $60$ & $60$\\
  \Xhline{2\arrayrulewidth} 
  Total        & $75$ & $719$ & $722$ & $722$ & $722$\\
  \Xhline{4\arrayrulewidth}
  \end{tabular}
  \caption{An overview of these selected corpora, including the number of actors and the distribution of utterances in the emotional classes.} 
  \label{tab:datasets}
\end{table}

\begin{table}[ht]
  \centering
  \begin{tabular}{lccccc}
  \Xhline{4\arrayrulewidth}
  Corpus & \multicolumn{4}{c}{Actor Set Partition} \\
  \cline{2-5}
  & 1 &  2 & 3 & 4 \\
  \Xhline{2\arrayrulewidth}
  RAVDESS     & $3$F, $3$M & $3$F, $3$M & $3$F, $3$M & $2$F, $3$M\\
  eNTERFACE   & $2$F, $9$M & $2$F, $8$M & $2$F, $8$M & $3$F, $8$M\\ 
  EMOVO       & $1$F, $0$M & $1$F, $1$M & $1$F, $1$M & $0$F, $1$M\\
  SAVEE       & $0$F, $1$M & $0$F, $1$M & $0$F, $1$M & $0$F, $1$M\\
  \Xhline{2\arrayrulewidth} 
  Total       & $6$F, $13$M & $6$F, $13$M & $6$F, $13$M & $5$F, $13$M\\
  \Xhline{4\arrayrulewidth}
  \end{tabular}
  \caption{F: female, M: male. The gender and corpus distributions in each actor set partition of the cross validation.} 
  \label{tab:cross-validation}
\end{table}

For the evaluation, we adopt a $4$-fold cross validation strategy. To begin with, we split the actor set into $4$ partitions. Moreover, we impose extra constraints to make sure that each partition is as gender and corpus uniform as possible. For example, each actor set partition is randomly distributed with $2$-$3$ female actors and $8$-$9$ male actors from the eNTERFACE'05 corpus. More details are provided in Table \ref{tab:cross-validation}. By partitioning the actor set, it becomes easier to maintain speaker independence between training and validation throughout all of the experiments.

\begin{table}[h]
  \centering
  \begin{tabular}{ccc}
  \Xhline{4\arrayrulewidth}
  Models & Layers & No. Params \\
  \Xhline{4\arrayrulewidth}
  $3$-Branch  & Conv2d(4,2,16) +      & 1.17 M \\
  	ResNet	    & BatchNorm2d + ReLU +   &      \\
  				    & (Shortcut, Branch$\times2$) +  &      \\
  $[$w/ shake reg.         & $[$ShakeShake $\{\times2\}\textrm{+}]$   &\\
  $\{$on \textbf{Both}$\}]$ & ReLU + Mean-Pooling + & \\
  				    & Dropout(0.5) + & \\
  				    & Linear(256) + ReLU +  & \\
  				    & Dropout(0.25) + & \\
  				    & Linear(256) + ReLU +  & \\
  				    & Linear(4) &\\
  \Xhline{2\arrayrulewidth} 
  Branch & Conv2d(4,2,64) + & 10.2 K\\
   				       & BatchNorm2d + ReLU + & \\
   				       & Conv2d(4,4,128) + & \\
   				       & BatchNorm2d & \\
  \Xhline{4\arrayrulewidth}
  \end{tabular}
  \caption{Network architecture, layers and the number of parameters in the baseline and proposed models. Conv2d$(N,H,W)$ stands for a $2D$ convolutional layer with $N$ filters of size $H\times W$ and Linear$(N)$ for a fully connected layer with $N$ nodes. The Mean-Pooling layer represents the temporal pooling for generating an utterance representation.} 
  \label{tab:model_spec}
\end{table}

\begin{table*}[ht]
  \centering
  \begin{tabular}{cccccccccccccccccccccc}
  \Xhline{4\arrayrulewidth}
  Model & \multicolumn{13}{c}{Patience in Early Stopping}\\
  \cline{2-14}
  &         
   9& 11& 13& 15& 17& 19& 21& 26& 31& 36& 41& 46& 51\\
  \Xhline{4\arrayrulewidth}
  \textbf{Baseline}&       
  48.01& 48.01& 48.57& 50.59& 51.78&  
  56.03& 56.03& 56.49& 57.66& 57.66&  
  57.66&  58.85&  58.85\\
  \Xhline{4\arrayrulewidth}
  \textbf{Full}&
  47.26& 47.26& 48.46& \underline{52.58}& \underline{\textbf{53.23}}&  
  53.23& 53.23& 56.24& 56.24& 56.94&  
  56.94&  57.34&  57.34\\

  \textbf{Upper}&
  47.30& \underline{48.62}& \underline{52.66}& \underline{53.31}& \underline{54.73}& 
  55.26& 55.47& 56.00& 57.50& 57.50& 
  \underline{57.79}& 57.79& 57.79\\

  \textbf{Lower}&
  45.62& 46.55& 47.66& 48.04& 48.79&  
  48.79& 49.21& 51.34& 51.48& 54.18&  
  54.18& 54.18& 54.18\\

  \textbf{Both}&
  46.97& \underline{49.66}& \underline{50.72}& \underline{51.61}& \underline{\textbf{54.13}}&  
  54.13& 54.13& 54.58& 55.08& 57.20&  
  \textbf{57.66}& 57.79& 57.79\\
  \Xhline{4\arrayrulewidth}
  \end{tabular}
  \caption{Averaged unweighted accuracy (\%) on the validation partition over $4$-fold cross validation.} 
  \label{tab:valid_ua}
\end{table*}

\begin{table*}[ht]
  \centering
  \begin{tabular}{cccccccccccccccccccccc}
  \Xhline{4\arrayrulewidth}
  Model & \multicolumn{13}{c}{Patience in Early Stopping}\\
  \cline{2-14}
  &         
   9& 11& 13& 15& 17& 19& 21& 26& 31& 36& 41& 46& 51\\
  \Xhline{4\arrayrulewidth}
  \textbf{Baseline}&       
  -0.14& -0.14& 1.45& 1.90& 4.50& 8.93&   
  8.93& 11.42& 14.74& 14.74& 14.74& 16.34& 16.34\\
  \Xhline{4\arrayrulewidth}
  \textbf{Full}&
  0.08& 0.08& \underline{0.84}& 3.16& \underline{\textbf{3.33}}& \underline{3.33}& \underline{3.33}& 
  \underline{8.05}& \underline{8.05}& \underline{10.57}& \underline{10.57}& \underline{11.13}& \underline{11.13}\\

  \textbf{Upper}&
  2.35& 2.68& 4.98& 5.85& 7.38& 10.47& 11.45& 
  14.09& 16.08& 16.08& 18.62& 18.62& 18.62\\

  \textbf{Lower}&
  0.13& \underline{-0.18}& \underline{0.84}& \underline{1.66}& \underline{3.34}& \underline{3.34}& \underline{3.99}&   
  \underline{8.56}& \underline{8.90}& \underline{14.27}& \underline{14.27}& \underline{14.27}& \underline{14.27}\\

  \textbf{Both}&
  1.41& 2.90& 2.74& 2.22& \underline{\textbf{2.73}}& \underline{2.73}& \underline{2.73}& \underline{4.57}&   
  \underline{6.13}& \underline{8.16}& \underline{\textbf{10.14}}& \underline{11.53}& \underline{11.53}\\
  \Xhline{4\arrayrulewidth}
  \end{tabular}
  \caption{Averaged gap between the unweighted accuracy (\%) on the training and validation partitions over $4$-fold cross validation.}
  \label{tab:gap_ua}
\end{table*}

\subsubsection{Experimental Setup}
\label{subsec:exp_setup}
To start with, we extract the spectrograms of each utterance with a $25$ms window for every $10$ms using the Kaldi \cite{Povey11thekaldi} library. Cepstral mean and variance normalization is then applied on the spectrogram frames per utterance. To equip each frame with a certain context, we splice it with $10$ frames in the left and $5$ frames in the right. Therefore, a resulting spliced frame has a resolution of $16\times257$. Since emotion involves a longer-term mental state transition, we further down-sample the frame rate by a factor of $8$ to simplify and expedite the training process. 

We establish a baseline of $3$-branch ResNet and list the details in Table \ref{tab:model_spec}. For each utterance, a simple mean pooling is taken at the output of the residual block to form an utterance representation before feeding it to the fully connected layers. We avoid explicit temporal modeling layers such as a long short-term memory recurrent network because our focus is on shaking the ResNet. Note that the ShakeShake layer has no parameter to learn and hence the model size does not change during this work. We implement the ShakeShake layer as well as the entire network architecture using the PyTorch \cite{pytorch} library. Only the Shake-Shake combination \cite{gastaldi2017} is used and shaking is applied independently per frame. Due to space limit, we leave other combinations for future work. The models are learned using the Adam optimizer \cite{KingmaB14} with an initial learning rate of $0.001$ and the training is carried out on an NVIDIA Tesla K80 GPU. We use a mini-batch of $64$ utterances across all model training and let each experiment run for $200$ epochs in order to investigate the regularization power when over-training occurs.

\subsubsection{Results}
\label{sec:results}
Table \ref{tab:valid_ua} and \ref{tab:gap_ua} summarize the benchmarking of the unweighted accuracy (UA) of cross validation and the gap of UA between training and validation with respect to different patience in early stopping. 

In Table \ref{tab:valid_ua}, the underlined numbers indicate when a model performs better than the baseline. A clear trend is that if the training process is stopped early, models with regularization tend to outperform the baseline. On the other hand, if the training goes too far, the situation is almost entirely the opposite. However, even when over-trained the margin that the baseline has over the other regularized models is only around $1\%$ except for the model \textbf{Lower}. One thing to note, in particular, is that the model \textbf{Lower} seems to struggle with difficulties in capturing the affective pattern since the beginning. 

In Table \ref{tab:gap_ua}, the underlined numbers indicate when a model has a smaller gap than that of a baseline under the same patience. The apparent trend here is that if we let the training keep going, almost all regularized models tend to have a smaller gap compared to the baseline; in other words, the baseline tends to overfit more under the same patience in early stopping. We also note the model \textbf{Upper}, despite being regularized, appears to have a larger gap than the baseline does since the beginning of learning.

Fortunately, these two trends overlap about when patience equals $17$. In both Table \ref{tab:valid_ua} and \ref{tab:gap_ua}, the boldfaced numbers represent when a model performs not worse than the baseline and has a smaller gap. Based on these two criteria, the models \textbf{Full} and \textbf{Both} both demonstrate a superior performance while staying far from being over-trained. Moreover, the model \textbf{Both} is able to match the performance of the baseline even in the over-trained region where patience equals $41$, while still achieving a smaller gap compared to the baseline. Another observation is that regularized models generally require more patience to reach the same gap, especially the model \textbf{Both}. This suggests early stopping under the same patience may not be an universally optimal strategy.

When benchmarked with the model \textbf{Full}, the model \textbf{Both} always has a higher accuracy whenever they achieve comparable gaps (e.g. $3.33$ versus $2.73$, $8.05$ versus $8.16$, etc), and most of the time when under the same patience. This phenomenon corroborates our hypothesis that independently shaking the sub-bands would help to learn a better model for affective computing. Nevertheless, the concerning fact that the models \textbf{Upper} and \textbf{Lower} show totally different characteristics requires further investigation in the future. 

Unfortunately, the improvement in Experiment I is not statistical significant, possibly due to a small degree of freedom ($\mathrm{df}=4-1=3$). The strong regularization by shaking may also suppress the expressiveness by the shallow architecture in Table \ref{tab:model_spec} in some folds. 

\subsection{Experiment II}
\label{subsub:expii}
To better demonstrate the effectiveness of shaking sub-bands, we present another set of experiments similar to Experiment I in this subsection, except that we employ two extra corpora and a deeper architecture here. Table \ref{tab:data_exp2} summarizes the corpora for Experiment II, including the additional IEMOCAP \cite{Busso2008} and EMO-DB \cite{BurkhardtPRSW05} corpora. A similar actor set partition is presented in Table \ref{tab:actor_set_exp2} to ensure speaker independence and gender and corpus-uniform distribution over each actor partition.

\subsubsection{Datasets}
\begin{table}[th]
  \centering
  \begin{tabular}{lccccc}
  \Xhline{4\arrayrulewidth}
  Corpus & No. & \multicolumn{4}{c}{No. Utterances}\\
  \cline{3-6}
  & Actors & joy & anger & sadness & fear \\
  \Xhline{2\arrayrulewidth} 
  eNTERFACE         & $42$ & $207$ & $210$ & $210$ & $210$ \\ 
  RAVDESS           & $24$ & $376$ & $376$ & $376$ & $376$\\
  IEMOCAP           & $10$ & $720$ & $1355$ & $1478$ & $0$ \\
  Emo-DB            & $10$ & $71$  & $127$  & $66$ &  $69$ \\
  EMOVO              & $6$  & $84$ & $84$ & $84$ & $84$\\
  SAVEE              & $4$  & $60$ & $60$ & $60$ & $60$\\
  \Xhline{2\arrayrulewidth} 
  Total        & $96$ & $1518$ & $2212$ & $2274$ & $799$\\
  \Xhline{4\arrayrulewidth}
  \end{tabular}
  \caption{$6803$ utterances} 
  \label{tab:data_exp2}
\end{table}

\begin{table}[h]
  \centering
  \begin{tabular}{lccccc}
  \Xhline{4\arrayrulewidth}
  Corpus & \multicolumn{4}{c}{Actor Set Partition} \\
  \cline{2-5}
  & 1 &  2 & 3 & 4 \\
  \Xhline{2\arrayrulewidth}
  eNTERFACE   & $2$F, $8$M & $2$F, $9$M & $3$F, $8$M & $2$F, $8$M\\ 
  RAVDESS     & $3$F, $3$M & $3$F, $3$M & $3$F, $3$M & $3$F, $3$M\\
  IEMOCAP     & $1$F, $2$M & $1$F, $1$M & $1$F, $1$M & $2$F, $1$M\\
  Emo-DB      & $1$F, $2$M & $1$F, $1$M & $1$F, $1$M & $2$F, $1$M\\
  EMOVO       & $1$F, $0$M & $1$F, $1$M & $1$F, $1$M & $0$F, $1$M\\
  SAVEE       & $0$F, $1$M & $0$F, $1$M & $0$F, $1$M & $0$F, $1$M\\
  \Xhline{2\arrayrulewidth} 
  Total       & $8$F, $16$M & $8$F, $16$M & $9$F, $15$M & $9$F, $15$M\\
  \Xhline{4\arrayrulewidth}
  \end{tabular}
  \caption{F: female, M: male.} 
  \label{tab:actor_set_exp2}
\end{table}

\begin{table}[h]
  \centering
  \begin{tabular}{cccc}
  \Xhline{4\arrayrulewidth}
  layer name & structure & stride & no.params \\
  \Xhline{4\arrayrulewidth}
  prelim-conv & $\left[2\times16, 4\right]\times 1$ & $[1,1]$ & $132$ \\
  \Xhline{\arrayrulewidth}
  res-8 & 
  $\begin{bmatrix}
  2\times16,  & 8  \\
  2\times16,  & 8  
  \end{bmatrix} \times 3$ & $[1,1]$ & $22.84$K\\
  \Xhline{\arrayrulewidth}
  res-16 & 
  $\begin{bmatrix}
  2\times16,  & 16  \\
  2\times16,  & 16  
  \end{bmatrix} \times 1$ & $[1,1]$ & $24.88$K\\
  \Xhline{\arrayrulewidth}
  res-32 & 
  $\begin{bmatrix}
  2\times16,  & 32  \\
  2\times16,  & 32  
  \end{bmatrix} \times 1$ & $[1,1]$ & $99.17$K\\
  \Xhline{\arrayrulewidth}
  average & - & - & - \\
  \Xhline{\arrayrulewidth}
  affine & $256\times 4$ & - & $1024$\\
  \Xhline{4\arrayrulewidth}
  total & -  & - & $148$K\\
  \Xhline{4\arrayrulewidth}
  \end{tabular}
  \label{tab:arch_exp2}
\end{table}

\begin{table}[h]
  \centering
  \begin{threeparttable}
  \begin{tabular}{ccc}
  \Xhline{6\arrayrulewidth}
  Model & Valid UA & Gap  \\
  \Xhline{4\arrayrulewidth}
  \textbf{Baseline} & $61.342$ & $7.485$ \\
  \Xhline{\arrayrulewidth}
  \textbf{Full} & $62.989$  & $-1.128$ \\
  \Xhline{\arrayrulewidth}
  \textbf{Both} & $\mathbf{64.973}$ & $\mathbf{1.791}$ \\
  \Xhline{6\arrayrulewidth}
  \end{tabular}
  \begin{tablenotes}\footnotesize
\item[*] average over three runs
  \end{tablenotes}
  \end{threeparttable}
  \label{tab:ua_gap_exp2}
\end{table}

\begin{table}[h]
  \centering
  \begin{threeparttable}
  \begin{tabular}{ccc}
  \Xhline{6\arrayrulewidth}
  Model Pair & Valid UA & Gap\\
  \Xhline{4\arrayrulewidth}
  \textbf{Baseline vs Full} & $1.44\times10^{-2}$ & $2.7\times10^{-4}$\\
  \Xhline{\arrayrulewidth}
  \textbf{Baseline vs Both} & $1.22\times10^{-5}$ & $3\times10^{-3}$ \\
  \Xhline{\arrayrulewidth}
  \textbf{Full vs Both} & $2.6\times10^{-3}$ & $9.9\times10^{-1}$ \\
  \Xhline{6\arrayrulewidth}
  \end{tabular}
  \caption{p-values resulted from one-sided paired t-test with df=$4\times3-1=11$}
  \end{threeparttable}
  \label{tab:p_value_exp2}
\end{table}

We employ a residual neural network with $3$ stages, res-8, res-16 and res-32, for Experiment II. There are $3$ residual blocks in res-8, and one residual block in res-16 and res-32. We remove the fully connected layer and keep only the output layer. Table \ref{tab:arch_exp2} contains details of the architecture.

We do not use early-stopping in Experiment II and let each model training run for $1000$ epochs, and for $3$ runs to obtain a robust estimate of the performance. Table \ref{tab:ua_gap_exp2} present the experimental results for UA and Gap. It is clear that both \textbf{Full} and \textbf{Both} are able to improve from \textbf{Baseline}, even with a much deeper architecture and with a longer training time.

Furthermore, we conduct statistical hypothesis testing to examine the significance of the observed improvement. The resulting p-values from one-sided paired t-test indicates that the improvements of \textbf{Full} and \textbf{Both} from \textbf{Baseline} are statistical significant, in terms of UA and Gap, in which \textbf{Both} has reached a higher UA but also a larger gap, compared to \textbf{Full}. In fact, the improvement of \textbf{Both} from \textbf{Full} is also significant in terms of UA. On the other hand, \textbf{Full} has significantly achieved a smaller gap, compared to \textbf{Both}. More details are presented in Table \ref{tab:p_value_exp2}.

\vspace{-1em}
\section{Conclusions}
\label{sec:conclude}
We have proposed a Shake-Shake regularized multi-branch ResNet model for speech emotion recognition. 
In particular, we have experimented with different configurations for the Shake-Shake regularization on the full band, the upper and the lower sub-bands alone and simultaneously. 
We conducted two experiments to study the shaking regularization mechanism for affective computing and demonstrated that sub-band shaking can lead to an improved UA and reduced overfitting.
The results support our hypothesis that shaking different sub-bands with independent strength would benefit learning in affective computing. 
With a commonly used patience, say $17$, the shallow models $\textbf{Full}$ and \textbf{Both} are able to deliver competitive performances over the baseline with reduced over-fitting. 
When experimenting with deeper models, we find that the shaking regularization keeps models from overfitting even with a substantially longer time.
The result from hypothesis testing shows not only the significance of the improvement from \textbf{Baseline} but also that different configurations of shaking regularization manifest a trade-off between expressiveness and compression.
However, given the opposite behaviors of the models \textbf{Upper} and \textbf{Lower}, further investigation is necessary in the future. 

\newpage
\bibliographystyle{IEEEbib}
\bibliography{strings,refs}

\end{document}